\documentclass{elektr}
\usepackage{hyperref}
\hypersetup{
colorlinks=true,
urlcolor=blue,
citecolor=blue}
\usepackage[all]{xy,xypic}
\usepackage{amsfonts,amssymb,amsmath,amsgen,amsopn,amsbsy,theorem,graphicx,epsfig}
\usepackage{eufrak,amscd,bezier,latexsym,mathrsfs,eurosym,enumerate}
\usepackage[utf8]{inputenc}
\usepackage[english]{babel}
\usepackage{cleveref,multirow}
\usepackage[dvipsnames]{xcolor}
\usepackage[pagewise]{lineno}
\usepackage{mathtools}

\usepackage{natbib}
\yil{}
\vol{}
\fpage{}
\lpage{}
\doi{}

\title{Capacity improvement of non-orthogonal multiple access downlink transmission by orbital angular momentum based mode division multiple access}

\author[Ahmed Al Amin and Soo Young Shin]{
\textbf{Ahmed Al Amin, Soo Young Shin\thanks{wdragon@kumoh.ac.kr}~}\\
IT Convergence Engineering, Kumoh National Institute of Technology, Gumi, South Korea %\\ ORCID iD: https://orcid.org/XXXX-XXXX-XXXX-XXXX\\

\\ [1.8em]

\rec{}
\acc{}
\finv{}
}

\def\E{\ifmmode{\mathbb E}\else{$\mathbb E$}\fi} %natural numbers
\def\N{\ifmmode{\mathbb N}\else{$\mathbb N$}\fi} %natural numbers
\def\R{\ifmmode{\mathbb R}\else{$\mathbb R$}\fi} %real numbers
\def\Q{\ifmmode{\mathbb Q}\else{$\mathbb Q$}\fi} %rational numbers
\def\C{\ifmmode{\mathbb C}\else{$\mathbb C$}\fi} %complex numbers
\def\H{\ifmmode{\mathbb H}\else{$\mathbb H$}\fi} %complex numbers
\def\Z{\ifmmode{\mathbb Z}\else{$\mathbb Z$}\fi} %integers
\def\P{\ifmmode{\mathbb P}\else{$\mathbb P$}\fi} %real numbers
\def\T{\ifmmode{\mathbb T}\else{$\mathbb T$}\fi} %real numbers
\def\SS{\ifmmode{\mathbb S}\else{$\mathbb S$}\fi} %real numbers
\def\DD{\ifmmode{\mathbb D}\else{$\mathbb D$}\fi} %real numbers

\newcommand{\bse}{\begin{subequations}}
\newcommand{\ese}{\end{subequations}}
\newcommand{\ben}{\begin{enumerate}}
\newcommand{\een}{\end{enumerate}}
\newcommand{\bens}{\begin{enumerate*}}
\newcommand{\eens}{\end{enumerate*}}
\newcommand{\be}{\begin{equation}}
\newcommand{\ee}{\end{equation}}
\newcommand{\bea}{\begin{eqnarray}}
\newcommand{\eea}{\end{eqnarray}}
\newcommand{\baa}{\begin{eqnarray*}}
\newcommand{\eaa}{\end{eqnarray*}}
\newcommand{\bc}{\begin{center}}
\newcommand{\ec}{\end{center}}

\theoremstyle{corollary}

\theoremstyle{lemma}

\theoremstyle{proposition}

\theoremstyle{axiom}

\theoremstyle{conjecture}

\theoremstyle{example}

\theoremstyle{definition}
\theoremstyle{remark}
\begin{document}

\maketitle

\begin{abstract}

In this paper, non-orthogonal multiple access (NOMA) downlink transmission is integrated with orbital angular momentum (OAM) based mode division multiple access (MDMA), called NOMA-OAM-MDMA. Different OAM modes can generate different OAM waves for different superimposed signals. So, every OAM wave will be transmitted a superimposed signal from the base station to the cell center user (CCU) and cell edge user (CEU). In this way, multiple OAM waves with multiple superimposed signals will transmit from BS to CCU and CEU simultaneously to enhance the capacity of NOMA downlink transmission. Finally, the effectiveness of the proposed schemes over the existing scheme and conventional orthogonal multiple access based scheme are demonstrated through the result analysis.

\end{abstract}

\section{Introduction}
\label{Int}

To cope with the large channel capacity requirements of future wireless networks, non-orthogonal multiple access (NOMA) has received significant attention from the research community [1-2]. NOMA is an interesting and emerging multiple access technique that is key player for the upcoming future wireless communication network because of its main fold spectral gains [3-4]. In a power domain NOMA (PD-NOMA) based system, multiple signals are superimposed in power domain at transmitter side, by utilizing the same code at the same frequency, which is not similar than existing orthogonal multiple access (OMA) schemes such as time division multiple access (TDMA), frequency division multiple access (FDMA) and code division multiple access (CDMA). At the receiver end, signal decoding techniques such as successive interference cancellation can be performed to decode each signal. Most of the recent research is mainly focus on the channel capacity improvement in the downlink (DL) transmission of NOMA [5-6]. Which inspires the research of this paper. Previous works proposed the use of user pairing [7], the combination of OMA and NOMA [8], integration of generalized space shift keying (GSSK) with NOMA [9], and the use of coordinated multiple points (CoMP) and NOMA [10]. All of these previous works are performed to improve the capacity of DL transmission of NOMA. 
There is a huge potential to utilize orbital angular momentum (OAM) signals to improve the channel capacities of the DL transmission of NOMA. OAM utilizes a new degree of freedom which is known as OAM mode for signal transmission [11-12]. OAM exploits the phase variation with respect to the azimuth angle of the propagated electromagnetic waves. This leads to the helical phase structure of the wave. A system model in [13] mode division multiple access using different OAM modes (OAM-MDMA) scheme exploits different OAM modes to boost the spectral efficiency. To improve the capacities of DL NOMA, a suitable technique is required which can improve the capacities of the DL transmission to the users as well as sum capacities (SC) to a great extent without any ICI.    

\par                      

In this paper, NOMA is integrated with OAM-MDMA. The name of the proposed NOMA with the OAM-MDMA scheme is NOMA-OAM-MDMA in this paper. Moreover, multiple waves are created by different OAM modes. The waves are exploiting to convey additional information to the cell center user (CCU) and cell edge user (CEU) of NOMA without ICI. In addition, by utilizing the active OAM modes to transport additional information to enhance the user capacities and sum capacities as well without any additional resources (e.g. time, frequency or power). Principle contributions of this paper are briefly described as follows: 

\begin{itemize}
\item NOMA is integrated with OAM-MDMA to improve the user capacities and SC as well. 

\item The capacities of CCU, CEU, and SC of the proposed NOMA-OAM-MDMA are analyzed and compared with conventional NOMA, and OMA-OAM-MDMA.    

\item The capacity improvement of the proposed scheme over existing schemes (e.g. NOMA) and OMA is manifested. As a benchmark, the user capacities and SC of OMA with OAM-MDMA (OMA-OAM-MDMA) are also compared with the capacities of the proposed scheme. 

\item The impact of normalized transmission distance between BS to CCU and number of OAM modes for NOMA-OAM-MDMA over user capacities and SC improvements of the proposed (NOMA-OAM-MDMA) scheme are analyzed and compared with conventional NOMA, and OMA-OAM-MDMA are also analyzed as well. 

\item By extensive computer simulations, the result analysis illustrates that the superiority of the proposed scheme over conventional NOMA, and OMA-OAM-MDMA schemes as well.  

\end{itemize}

The rest of this paper is organized as follows. Section II describes the related works.
Section III describes the system model and the proposed scheme explicitly. 
Section IV exhibits numerical result analysis. This paper is concluded in Section V.

\section{Related Works} 

Different recent research has been done to improve the spectral efficiency of NOMA users and SC. Spatial modulation (SM) with NOMA provides improvement of channel capacities in [14]. cooperative relaying strategy is a feasible strategy to enhance the SC of the NOMA DL transmission  [15-16]. Efficient resource management is another strategy  to enhance the sum capacity of NOMA [17]. Moreover, a suitable solution is required to improve the channel capacities of the users and SC of NOMA DL transmission without ICI as well. So an OAM based solution can overcome the challenges of NOMA DL transmission.  

OAM is a relatively new wireless communication paradigm that is attracting more and more attention for the potential short-distance line of sight (LOS) applications. Such as high-bandwidth backhaul communication for cellular networks [18]. Besides from conventional radio frequency-based communication, OAM has been extensively explored form free-space optical and fiber optic based communication system [19-20]. Moreover, mode combination in an ideal wireless OAM with Multiple Input Multiple Output (OAM-MIMO) multiplexing system can enhance the capacity of the system [21]. Which can provides better results instead of LOS-MIMO for cellular backhaul communication networks.  Furthermore, [22] shows that OAM with spatial modulation provides higher capacity improvement than MIMO in case of millimeter wave communication.     

\section{System Model}

A NOMA-OAM-MDMA network with a base station (BS) and two users (a cell-center user (CCU) and a cell-edge user (CEU)) is considered. The BS directly communicates with the CCU called as $UE_1$ and CEU called as $UE_2$ simultaneously by utilizing different signals between BS and respected users. In this case (NOMA-OAM-MDMA), total $N$ number of OAM modes can be generated. The signals are created by utilizing a set of activated OAM modes {$L$}, where $L=\{0,1... N-1\}$ is the set of activated OAM modes for the classical OAM-MDMA [13]. The same wave by utilizing the same OAM mode $l$ is conveying information from BS to $UE_1$ and $UE_2$ simultaneously, where $l \epsilon L$. The proposed network model is illustrated in Figure 1. The different waves by different OAM modes transmit a different superimposed signal to the $UE_1$ and $UE_2$. Moreover, Two aligned uniform circular arrays (UCA) facing each other and consisting of $N$ elements for the transmission and reception between the BS and the users [23]. Hereafter, subscript s,1, and 2 denote BS, $UE_1$, and $UE_2$, respectively. The distance between BS to $UE_1$ is $d_{s,1}$ and the distance between BS to $UE_2$ is $d_{s,2}$. the radii of the transmit and receive UCAs are given by $r_{tx_i}$ and $r_{rx_j}$ respectively [23]. The channel coefficient $h_{i,j}$ is the free space channel coefficient between any two nodes due to line-of-sight (LOS) communication [23]. Whereas , $(i,j \epsilon \{s, UE_1, and \ UE_2\})$. The LOS channel with additive white gaussian noise (AWGN) is considered for all cases in this paper because OAM is performed only for LOS communication cases [13,21,23]. The data communication policy along with the signal-to-interference-plus-noise ratio (SINR) model of the proposed scheme is discussed in detail in section 3.1.

\begin{figure}[h!]
\centering
\includegraphics[width=0.8\textwidth]{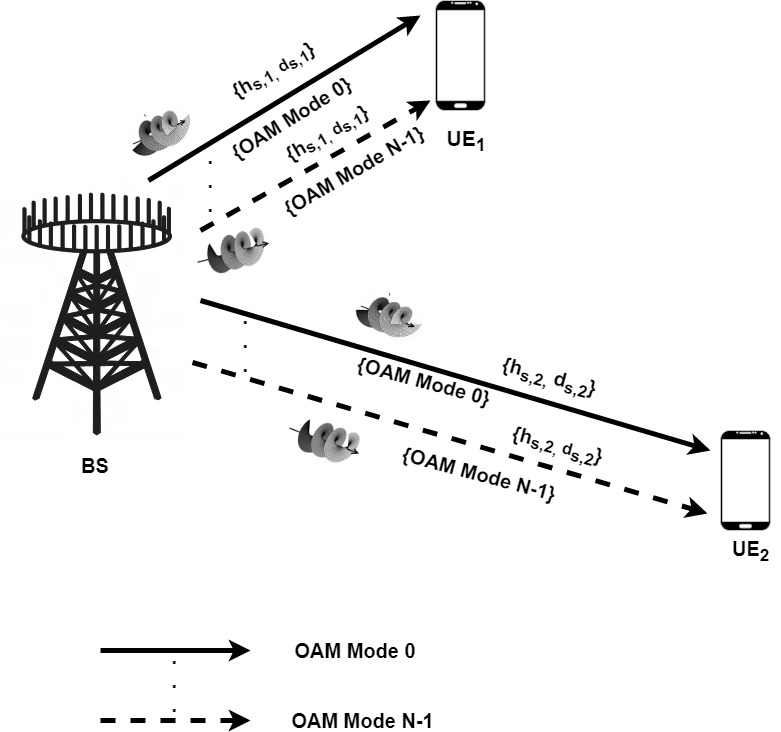}
\caption{Proposed system model for NOMA-OAM-MDMA}
\label{image-myimage}
\end{figure}
    
\subsection{NOMA-OAM-MDMA}

Following the principle of NOMA, a superimposed signal composite signal $A_l=\sqrt{p_{1_l}P_l}x_{1_l}+\sqrt{p_{2_l}P_l}x_{2_l}$ for $l$ mode of OAM wave, where $x_{1_l}$, $x_{2_l}$, and $p_{1_l}$, $p_{2_l}$ are the data symbols and the power allocation factors, respectively. Note that $x_{1_l}$, $p_{1_l}$ and $x_{2_l}$, $p_{2_l}$ are assigned to $UE_1$ and $UE_2$, respectively [13,23]. In case of NOMA, $p_{1_l}+p_{2_l}=P_l$ and $p_{1_l}<p_{2_l}$. Whereas, $P=\sum_{l=0}^{N-1} P_l^2$ is the total transmit power for downlink transmission. $T$ is the total duration of downlink transmission. Moreover, the total power allocation factor $P_l$ for each OAM mode $l$ from BS to the users are given below, 

\begin{equation}
P_{l}=\sqrt{\frac{\frac{P}{|\xi_l|^2}}{\frac{1}{|\xi_0|^2}+\frac{1}{|\xi_l|^2}+....+\frac{1}{|\xi_{N-1}|^2}}} ,
\end{equation}

To radiate an OAM wave with mode number $l$, the same input signal $A_l$ should be applied to the antenna element in UCA with a successive phase shift from element to the element of BS. Hence, the signal applied to the $n^{th}$ element of BS [13, 23], where the eigenvalues of the free space LOS channels are $\xi_0,\xi_1, ..... \xi_{N-1}$. Consequently, the superimposed signal applied to the $n^{th}$ element in case of is given for $n=0,...N-1$ as  

\begin{equation}
s^l_n=\frac{1}{\sqrt{N}} (A_l) e^{(-j2\pi (nl)/N)},
\end{equation}

With MDMA concept of OAM, the antenna element of BS $n \epsilon \{ 0,1,....N-1\}$ is fed by the linear superposition of the signals of different modes as [23],

\begin{equation}
s^n=\frac{1}{\sqrt{N}} \sum_{l \epsilon L} (A_l) e^{(-j2\pi (nl)/N)},
\end{equation}

where $n=0,1,....N-1$. For LOS propagation environment and AWGN for $UE_1$ and $UE_2$, the free space channel response between BS to $UE_1$ and $UE_2$ are as below for OAM mode $l$ [13,23],

\begin{equation}
h_{{i,j}_l}=\beta \frac{\lambda}{4 \pi d_{i,j}} e^{j k d_{n_i,p_j}},
\end{equation}

where, $k=\frac{2 \pi}{\lambda}$, $\lambda$ is the wavelength, $\beta$ is a constant related with antenna gains, and in case of perfect alignment between BS and users UCAs $d_{n_i,p_j}=\sqrt{d_{i,j}^2+r_{tx_i}^2+r_{rx_j}^2-2r_{tx}r_{rx}\phi_{n_i,p_j}}$. Moreover, $\phi_{n_i,p_j}=\frac{2\pi (n_i-p_j)}{N}$ and it should be noted that due to the employment of UCAs at both BS and user sides as well as the construction of a configuration with perfect symmetry [23]. Moreover, $n_1 \sim CN(0,\sigma^2)$ and $n_2 \sim CN(0,\sigma^2)$ are the complex AWGN at $UE_1$ and $UE_2$ respectively with zero mean and variance $\sigma^2$. Moreover, $|h_{s,1}|^2>|h_{s,2}|^2$ because $d_{s,1}<d_{s,2}$ is considered in this paper. The demultiplexed and recovered receive signal at $UE_1$ and $UE_2$ are respectively given by [13,23]

\begin{equation}
y_{1_l}= \frac{1}{\sqrt{N}} \sum_{p_1=0}^{N-1} (y_{1_{p_1}}) e^{(-j2\pi (p_1 l)/N)} + n_1,
\end{equation}

\begin{equation}
y_{2_l}= \frac{1}{\sqrt{N}} \sum_{p_2=0}^{N-1} (y_{l_{p_2}}) e^{(-j2\pi (p_2 l)/N)} +n_2,
\end{equation}

for $l=0,1,....N-1$, where $y_{1_l}$ and $y_{2_l}$ are the $p_1$th or $p_2$th element of received signal by $UE_1$ and $UE_2$ respectively. According to the downlink NOMA protocol, $UE_1$ first decodes $x_2$ and then performs SIC to decode own symbol $x_1$. Thus, the received SINR at $UE_1$ in case of $x_1$ and $x_2$ are respectively given by following equations due to received OAM wave with $l$ mode,  

\begin{equation}
\gamma_{x_{1_l}}^{UE_1}={\rho_l{{|h_{s,1_l}|}^2}p_{1}},
\end{equation}
\begin{equation}
\gamma_{x_{2_l}}^{UE_1}=\frac{\rho_l{{|h_{s,1_l}|}^2}p_{2}}{\rho_l{{|h_{s,1_l}|}^2}p_{1}+1},
\end{equation}

where $\rho_l \triangleq \frac{P_l}{\sigma^2}$ is the transmit signal-to-noise ratio (SNR) by BS for OAM mode $l$ and total transmit SNR, $\rho=\sum_{l=0}^{N-1} \rho_l$.  The signal can be directly decoded at $UE_2$ by treating the signal $x_1$ as noise. Therefore, the received SINR at $UE_2$ in case of  $x_2$ is given by following equations due to received OAM wave with $l$ mode,

\begin{equation}
\gamma_{x_{2_l}}^{UE_2}=\frac{\rho_l{{|h_{s,2_l}|}^2}p_{2}}{\rho_l{{|h_{s,2_l}|}^2}p_{1}+1},
\end{equation}

\subsection{Achievable capacity analysis}
By considering normalized total time duration $T=1$ for downlink transmission. The equations for the capacities are derived in the following segments. 

\subsubsection{Capacity of $UE_1$}
$x_{1_l}$ and $x_{2_l}$ are received by $UE_1$ for OAM mode $l$. So, the achievable capacity of $UE_1$ for the wave with OAM mode $l$ is obtained as below by Eq. 6,

\begin{equation}
C_{UE_{1_l}}=\log_2(1+{\mu_{1,l}}\gamma_{x_{1_l}}^{UE_1}),
\end{equation}

Where, $\mu_{1,l}$ is the singular value of the channel response matrix $h_{s,1_l}$ for OAM mode $l$. So, the total channel capacity at $UE_1$ for all the OAM wave from BS to $UE_1$ is as below,

\begin{equation}
C_{UE_1}=\sum_{l={0}}^{N-1}\log_2(1+{\mu_{1,l}}\gamma_{x_{1_l}}^{UE_1}),
\end{equation}

\subsubsection{Capacity of $UE_2$}
$x_{2_l}$ is directly decoded by $UE_2$ by treating $x_{l_1}$ as noise. So, the achievable capacity of $UE_2$ for the wave with OAM mode $l$ is obtained as below by Eq. 8,

\begin{equation}
C_{UE_{2_l}}=\log_2(1+{\mu_{2,l}}\gamma_{x_{2_l}}^{UE_2}),
\end{equation}

Where, $\mu_{2,l}$ is the singular value of the channel response matrix $h_{s,2_l}$ for OAM mode $l$. So, the total channel capacity at $UE_2$ for all the OAM wave from BS to $UE_2$ is as below,

\begin{equation}
C_{UE_2}=\sum_{l={0}}^{N-1}\log_2(1+{\mu_{2,l}}\gamma_{x_{2_l}}^{UE_2}),
\end{equation} 

\subsubsection{Sum Capacity}

So, the SC can be achieved by adding Eq. 11 and Eq. 13 as below,
\begin{equation}
C_{SC}=C_{UE_1}+C_{UE_2}
\end{equation}

\subsection{OMA-OAM-MDMA}

For a fair comparison with the proposed NOMA-OAM-MDMA scheme, the OMA-OAM-MDMA scheme is also devised in this paper as a benchmark, taking into account time-division multiple access. It should be mentioned that eight time slots are required for completing data transmission in OMA-OAM-MDMA, whereas only one time slot is required in the proposed NOMA-OAM-MDMA scheme. The capacities of the users and the SC are given by

\begin{equation}
C_{UE_1}^{OMA}=\frac{1}{8}\sum_{l=0}^{N-1} \log_2(1+{\mu_{1,l}\rho_l{|h_{s,1_l}|}^2}).
\end{equation}

 \begin{equation}
C_{UE_2}^{OMA}=\frac{1}{8}\sum_{l=0}^{N-1}\log_2(1+{\mu_{2,l}\rho_l{|h_{s,2_l}|}^2}).
\end{equation}

So, the SC can be achieved by adding Eq. 15 and Eq. 16 as below,
\begin{equation}
C_{SC}^{OMA}=C_{UE_1}^{OMA}+C_{UE_2}^{OMA}.
\end{equation}

\subsection{Numerical Results}

In this section computer simulation results for the user capacities and SC of the proposed scheme are examined as well as explained. The impact of changes in transmit SNR $\rho$, normalized transmission distance between BS and CCU $\tilde{{d_{s,1}}}
$ are discussed. Collinear placement of all nodes (e.g. BS, $UE_1$, and $UE_2$ and normalized distances between any two nodes are considered, where $d_{s,1}=500 \lambda$, $d_{s,2}= 1000 \lambda$. The wave length $\lambda=0.03$, $T=1$, $L=\{0,1,2,3\}$ (For 4 modes of OAM), $L=\{0,1\}$ (For 2 modes of OAM)and $P=1$ are considered as well for simulation purpose. For performance comparison, simulation results for CCU capacity ($UE_1$ capacity), CEU capacity ($UE_2$ capacity) and SC of NOMA-OAM-MDMA, conventional NOMA and OMA-OAM-MDMA are also provided. In the ideal case, each receiver is located in the OAM circle region of OAM signals with required OAM mode by utilizing the directional characteristics of OAM beam. Note that, similar simulation parameters are considered for the proposed and compared schemes for consistency.    

Impact of $\rho$, $d_{s,1}$ and $N$ on CCU capacity, CEU capacity and SC of the proposed system is shown in this part. All figures are plotted for different LOS channels from BS to CCU and CEU for different OAM signals.  

\begin{figure}[h!]
\centering
\includegraphics[width=0.9\textwidth]{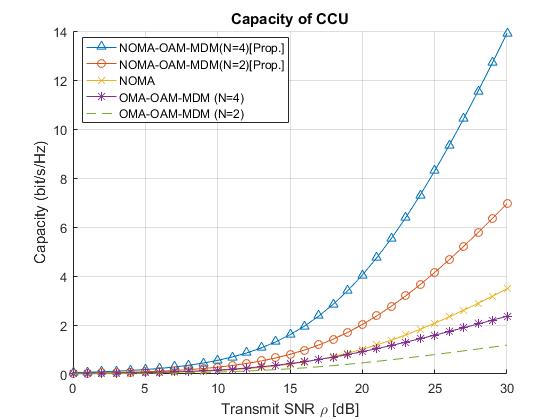}
\caption{Capacity comparisons of CCU with respect to transmit SNR $p_1=0.4$, $p_2=0.6$, $P=1$, $T=1$, $\lambda=0.03$, $d_{s,1}=500 \lambda$, and $d_{s,2}=1000 \lambda$.}
\label{image-myimage}
\end{figure}
\par

The capacity of CCU ($UE_1$) with respect to (w.r.t.) transmit SNR $\rho$ is demonstrated in Figure 2 for the proposed NOMA-OAM-MDMA($N=4$ and $N=2$) and compared with existing NOMA, and OMA-OAM-MDMA ($N=4$ and $N=2$). Parameters $p_1=0.4$, and $p_2=0.6$ are set during the simulation. The CCU capacity of all schemes increases linearly with an increase of $\rho$. The proposed scheme (NOMA-OAM-MDMA) exhibits far better performance than conventional NOMA and other OMA-OAM-MDMA schemes. Moreover, larger set of OAM modes $L=\{0,1,2,3 \}$ due to $N=4$ for the proposed scheme provides much better capacity for CCU than lower set of OAM signals $L=\{0,1\}$ due to $N=2$. Because of the higher number of OAM signals carrying much more superimposed signals than a lower number of signals without ICI. This is performed by utlizing OAM based MDMA for different OAM signals. The OAM signals are generated by each different OAM modes. As a result, the capacity of CCU is enhanced significantly compared to other schemes. The same phenomena happened to the OMA-OAM-MDMA case as well. A higher number of signals provides much better capacity than the lower number of OAM signals. Moreover, due to utilizing separate time slot for each transmission the CCU capacity of OMA-OAM-MDMA schemes are far lower than the proposed scheme with the higher number of OAM signals which is illustrated in Figure 2 as well.

\begin{figure}[h!]
\centering
\includegraphics[width=0.8\textwidth]{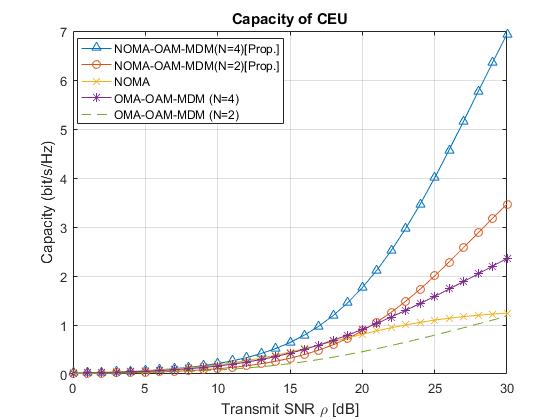}
\caption{Capacity comparisons of CEU with respect to transmit SNR $p_1=0.4$, $p_2=0.6$, $P=1$, $T=1$, $\lambda=0.03$, $d_{s,1}=500 \lambda$, and $d_{s,2}=1000 \lambda$.}
\label{image-myimage}
\end{figure}
\par

The capacity of CEU ($UE_2$) with respect to (w.r.t.) transmit SNR $\rho$ is demonstrated in Figure 3 for the proposed NOMA-OAM-MDMA($N=4$ and $N=2$) and compared with existing NOMA, and OMA-OAM-MDMA ($N=4$ and $N=2$). Parameters $p_1=0.4$, and $p_2=0.6$ are set during the simulation. The CEU capacity of all schemes increases linearly with an increase of $\rho$. The proposed scheme (NOMA-OAM-MDMA) exhibits far better performance than conventional NOMA and other OMA-OAM-MDMA schemes. Moreover, higher set of OAM modes $L=\{0,1,2,3\}$ due to $N=4$ for the proposed scheme provides much better capacity for CCU than lower set of OAM modes $L=\{0,1\}$ due to ($N=2$). Because of the higher number of signals carrying much more superimposed signals than a lower number of signals without ICI like as CCU. This is performed by utlizing OAM based MDMA for different OAM signals. The OAM signals are generated by each different OAM modes. As a result, the capacity of CEU is enhanced significantly compared to other schemes. The same phenomena happened to the OMA-OAM-MDMA case as well. A higher number of signals provides much better capacity than the lower number of OAM signals. Moreover, due to utilizing separate time slot for each transmission the CEU capacity of OMA-OAM-MDMA schemes are far lower than the proposed scheme with the higher number of OAM signals which is illustrated also in Figure 3.

\begin{figure}[h!]
\centering
\includegraphics[width=0.8\textwidth]{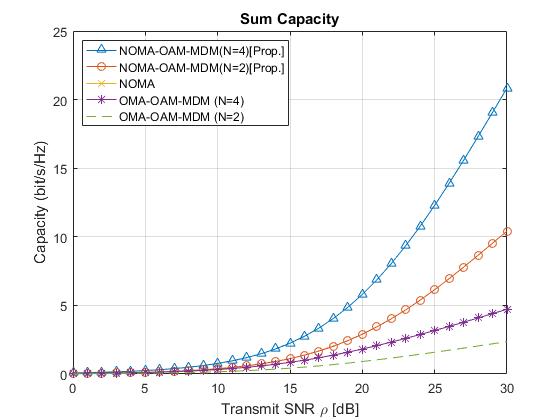}
\caption{SC with respect to transmit SNR $p_1=0.4$, $p_2=0.6$, $P=1$, $T=1$, 
$\lambda=0.03$, $d_{s,1}=500 \lambda$, and $d_{s,2}=1000 \lambda$.}
\label{image-myimage}
\end{figure}
\par

In Figure 4, proposed NOMA-OAM-MDMA provides significantly higher SC than other compared schemes. Parameters $p_1=0.4$, and $p_2=0.6$ are set during the simulation. Moreover, higher $N$ provides better SC than lower $N$ for the proposed scheme as well. These phenomena happen because the CCU and CEU capacity is higher for the proposed scheme. Moreover, the higher number of $L$ due to the higher number of $N$ carrying much more superimposed signals to the CCU and CEU by utilizing OAM-MDMA for different modes of OAM as well. Hence, the proposed NOMA-OAM-MDMA scheme with a higher $N=4$ provides significantly higher SC than other compared schemes as well.

\begin{figure}[h!]
\centering
\includegraphics[width=0.8\textwidth]{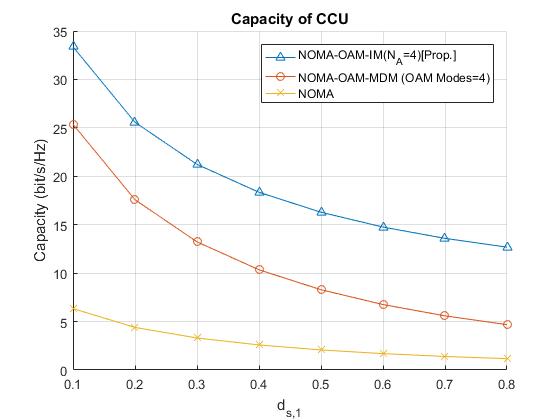}
\caption{CCU Capacity with respect to $d_{s,1}$ $p_1=0.4$, $p_2=0.6$, $P=1$, $T=1$, $\lambda=0.03$, $\rho=25 dB$ }
\label{image-myimage}
\end{figure}
\par

The impact of normalized transmission distance between BS to CCU $\tilde{{d_{s,1}}}
$ on CCU capacity is illustrates in Figure 5, where $\tilde{{d_{s,1}}}=\frac{{d_{s,1}}}{{d_{s,2}}}$. Parameters $p_1=0.4$, $p_2=0.6$, and $\rho=25 dB$ are set during the simulation. The CCU capacity is decreasing for increasing values of $\tilde{{d_{s,1}}}
$. The proposed scheme NOMA-OAM-MDMA provides better capacity for CCU due to variation of $\tilde{{d_{s,1}}}$. Moreover, higher values of $N$ for the proposed scheme provides significantly higher CCU channel capacity than other compared scheme. This enhancement is achieve at CCU  because higher number of signals (For $N=4$ provides $L=\{0,1,2,3 \}$ and For $N=2$ provides $L=\{0,1 \}$) conveying higher number of superimposed signals to the CCU. So the CCU capacity is enhanced for the proposed scheme with higher number of $N$ for the proposed scheme.

\begin{figure}[h!]
\centering
\includegraphics[width=0.8\textwidth]{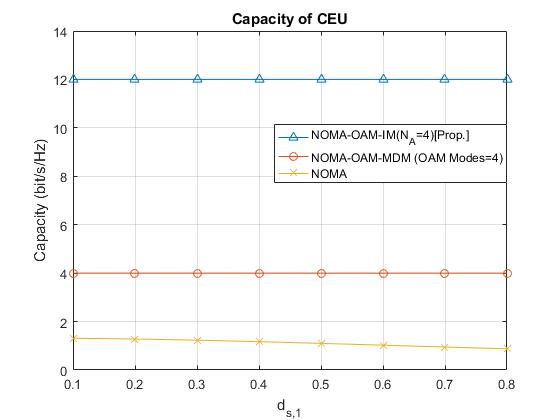}
\caption{CEU Capacity with respect to $d_{s,1}$ $p_1=0.4$, $p_2=0.6$, $P=1$, $T=1$, $\lambda=0.03$, $\rho=25 dB$ }
\label{image-myimage}
\end{figure}
\par

The impact of normalized transmission distance between BS to CCU $\tilde{{d_{s,1}}}
$ on CEU capacity is illustrates in Figure 6. Parameters $p_1=0.4$, $p_2=0.6$, and $\rho=25 dB$ are set during the simulation. In this analysis the position of CCU is varying but the position of CEU is remain steady. Hence the CEU capacity is steady for increasing values of $\tilde{{d_{s,1}}}$. Because the transmission distance between BS and CEU $d_{s,2}$ and $\rho$ are remain constant in this case. Moreover, the proposed scheme with higher $N$ outplayed other compared schemes in this observation as well. This enhancement is achieve at CEU  because higher number of signals (For $N=4$ provides $L=\{0,1,2,3\}$ and For $N=2$ provides $L=\{0,1 \}$) conveying higher number of superimposed signals to the CEU. So, the CEU capacity is enhanced for the proposed scheme with higher number of $N$ for the proposed NOMA-OAM-MDMA scheme.

\begin{figure}[h!]
\centering
\includegraphics[width=0.8\textwidth]{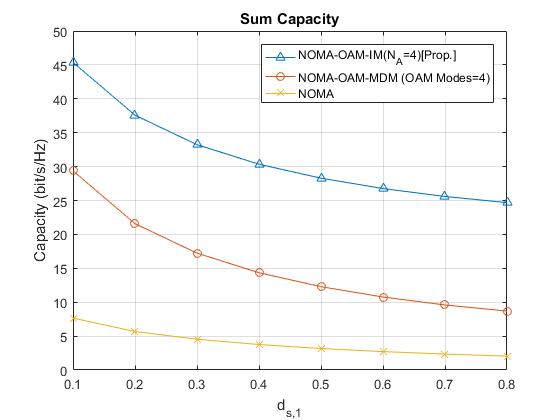}
\caption{SC with respect to $d_{s,1}$ $p_1=0.4$, $p_2=0.6$, $P=1$, $T=1$, $\lambda=0.03$, $\rho=25 dB$ }
\label{image-myimage}
\end{figure}
\par

The impact of normalized transmission distance between BS to CCU $\tilde{{d_{s,1}}}
$ on SC is illustrates in Figure 7. Parameters $p_1=0.4$, $p_F=0.6$, and $\rho=25 dB$ are set during the simulation. The SC is decreasing for increasing values of $\tilde{{d_{s,1}}}$ because the CCU capacity is decreasing for increasing values of $\tilde{{d_{s,1}}}$. This phenomena is shown in Figure 5. Moreover, the CEU capacity is steady for the proposed scheme and others as well. In addition, the proposed NOMA-OAM-MDMA scheme provides significantly SC than other compared schemes for higher $N$. Because higher values of $N$ provides higher CCU and CEU capacities which is shown in Figure 5 and Figure 6. As a matter of fact the SC is also improved as well for the proposed scheme with higher $N$ compared to NOMA-OAM-MDMA ($N=2$), conventional NOMA, OMA-OAM-MDMA ($N=4$), and OMA-OAM-MDMA ($N=2$).

\section{Conclusion}

In this paper, the NOMA-OAM-MDMA scheme has been proposed to enhance the user capacities and SC of NOMA downlink transmission. Moreover, conventional NOMA and OMA-OAM-MDMA are also compared with the proposed scheme for a fair comparison. According to result analysis, it is shown that the proposed scheme with the higher number of OAM signals provides improved performance compared to other conventional schemes in terms of capacities. In the future, the work can be extended by incorporating relay assisted CNOMA with the proposed scheme.

\section*{Acknowledgment}
This work was supported by the National Research Foundation of Korea(NRF) grant funded by the Korea government(MEST) (No. 2019R1A2C1089542) 
\bibliographystyle{plain}
%\bibliography{reference.bib}

\end{document}